\documentclass[11pt]{book}
\usepackage{Files/Wiley-AuthoringTemplate}
\usepackage{chapterbib}
\usepackage{longtable}
\usepackage{subcaption}
\usepackage[usestackEOL]{stackengine}
\usepackage[sectionbib,numbers, sort]{natbib}
\usepackage{tikz}
\renewcommand\thesection{\arabic{section}}
\makeindex

\begin{document}

\title{\Huge Joint secure communication and sensing in 6G networks
}

\author*[1]{Miroslav Mitev}
\author[1]{Amitha Mayya}
\author[1,2]{Arsenia Chorti}

\address[1]{\orgdiv{Wireless connectivity group}, 
\orgname{Barkhausen Institut}, 
\postcode{01187}, 
     \city{Dresden},
     \country{Germany}}%
     
\address[2]{\orgdiv{ETIS}, 
\orgname{UMR 8051 CY Cergy Paris Université, ENSEA, CNRS}, 
\postcode{95000}, 
     \city{Cergy}, 
     \country{France}}%


\address*{Corresponding Author:  \email{miroslav.mitev@barkhauseninstitut.org}}
\maketitle









Joint communication and sensing is expected to be one of the  features introduced by the sixth-generation (6G) wireless systems. This will enable a huge variety of new applications, hence, it is important to find suitable approaches to secure the exchanged information. Conventional security mechanisms may not be able to meet the stringent delay, power, and complexity requirements which opens the challenge of finding new lightweight security solutions. A promising approach coming from the physical layer is the secret key generation (SKG) from channel fading. While SKG has been investigated for several decades, practical implementations of its full protocol are still scarce. The aim of this chapter is to evaluate the SKG rates in real-life setups under a set of different scenarios. We consider a typical radar waveform and present a full implementation of the SKG protocol. Each step is evaluated to demonstrate that generating keys from the physical layer can be a viable solution for future networks. However, we show that there is not a single solution that can be generalized for all cases, instead, parameters should be chosen according to the context.


\section{Introduction} \label{sec:intro}

The development of wireless communications and radar sensing technologies has been progressing independently for many years. Their design has been fully separated, relying on different resources (frequency, hardware, waveform, etc.). However, as both services depend on the same principle, i.e., transmitting/receiving electromagnetic signals, it is envisioned that a co-designed approach can bring a number of benefits. In fact, the idea of a joint design is seeing a growing interest by both industry and academia and it is expected to become a key enabling technology for variety of 6G applications~\cite{joint_desing_JCAS}. 

A simple approach to integrate both services in one device is by sharing the same spectrum but transmitting separate signals. The main problem for this case is finding a flexible interference cancellation technique that can smoothly operate in different scenarios~\cite{enabling_JCAS}. Another, more efficient, approach is to use a single signal for both communication and sensing. This approach can provide higher spectral efficiency but requires joint waveform design. In this sense, depending on the application the waveform design can be communication-centric (suitable for application with high data rate requirements and low sensing precision, e.g., smart home applications) and radar-centric (suitable for application with low data rate requirements and high sensing precision, e.g., vehicular networks).

Communication signals have complex structure and are typically modulated in one or multiple domains, e.g., time, spatial and frequency. This increases the cost as advanced transmitter/receiver modules~\cite{enabling_JCAS} are required. On the other hand, signals used for radar have a simple structure, e.g., pulsed or continuous waves. Such waveforms are specifically designed for sensing and do not require the deployment of advanced processing modules. Such signals cannot provide high throughput, however, their achievable data rates can still be sufficient for certain sensing-oriented applications~\cite{Andre_chirp_eucap21}. Some of the key performance indicators for such applications are high sensing precision, low-latency, low-complexity and power efficiency~\cite{dual_function_JCAS}. 

A question that naturally arises is: \textit{How can we secure the communication in radar-centric applications?} It is clear that current cryptographic algorithms may not be able to support the requirements listed above. As an example, \cite{Teniou18} has reported that the verification of digital signature over $400$ MHz processor in a vehicular to everything (V2X) network exceeds the low-latency requirements and requires approximately $20$ ms. Furthermore, traditional asymmetric public-key encryption (PKE) protocols do not satisfy low-energy requirements and cannot scale well for massive deployment~\cite{ZHANG}. Finally, the advances in quantum computing can leave current security solutions vulnerable, unless key sizes increase significantly~\cite{MOSCA}. Therefore, proposals for new lightweight security alternatives are sought.   

Following from the above, the goal of this chapter is to provide a performance evaluation of one lightweight and quantum-secure key distribution technique (currently considered for 6G~\cite{6g_white_paper}). In particular, we investigate the physical layer security (PLS)-based secret key generation (SKG) with a focus on sensing-oriented networks. The SKG process (introduced in \cite{Maurer} and \cite{Csiszar}) allows two devices to extract a common random key sequence relying on the stochastic behaviour and reciprocity of wireless channels. Some of the advantages of SKG are that, it does not require a complex infrastructure (as in the case of PKE), it can be employed using an arbitrary PHY waveform (if the transmitted  power spectral density (PSD) is known) and it does not require matched receivers~\cite{Zoli_EURASIP2020, ISWCS_2021}. This work focuses on an SKG approach that utilizes one of the widely used signals for radar sensing, i.e., the compressed high resolution pulse (chirp).

Thanks to their wide bandwidth, chirps can be used to harvest entropy from the wireless channel. To extract channel observations we rely on a filterbank-based approach, where keys are generated from power measurements at different frequencies. This approach has been proposed in~\cite{Zoli_EURASIP2020} and analytically/numerically analyzed by some of the authors of this chapter in~\cite{ISWCS_2021, MITEV_globecom2022, MITEV_vtc2022}. As a continuation of these works, here, we take a step further and evaluate the performance of the approach in an experimental setup. A measurement campaign is performed in an indoor environment considering a set of different scenarios. To arrive at the final key rate all steps of the SKG protocol are implemented. First, advantage distillation is performed by applying a filterbank over received signals. Next, information reconciliation is done using Polar codes. To evaluate the key generation rate we pass the output of the reconciliation through a conditional min-entropy (conditioning on an eavesdropper knowledge)  estimation.

The rest of this chapter is organized as follows: Section \ref{sec:motivation} motivates this work, Section \ref{sec:sys_model} gives an overview of our system model, Section \ref{sec:SKG} gives a detailed overview of the SKG protocol, Section \ref{sec:measurement_campaign}, presents the details of the measurement campaign, Section \ref{sec:results} summarizes some of the main results of this work and concludes this chapter.

\section{Related work and motivation}\label{sec:motivation}

Radar systems are already deployed in numerous verticals including, automotive industry, ship and aircraft navigation, smart factories and many more~\cite{jcas_road_ahead}. It is clear that the applications above can benefit from a co-designed communication and sensing approach. Particularly, vehicles can continuously sense the environment while sharing the obtained information with others on the road~\cite{Andre_chirp_SKG_2020}. Similarly, autonomous machinery can communicate to improve production, while sensing for approaching people. Furthermore, in bistatic (or multistatic) radar applications, radars can accurately detect multiple targets and simultaneously convey location information to improve safety, e.g., in traffic control. Generally, in many joint communication and sensing scenarios, the data to be transmitted consists of confidential information, including location, speed and identity. Therefore, it is important to protect the transmission from possible information leakage and provide confidentiality.


As noted in the 3GPP report~\cite{3gppURLLC} a major limitation of 5G networks, especially in the context Internet of things (IoT), is related to security. Standard cryptographic solutions rely on computationally intensive modulo arithmetic operations which makes them unsuitable for power constraint devices. This issue has also been long investigated by the national institute of standards and technology (NIST). Particularly an investigation on  cryptography for constrained environments was initiated by NIST in 2013~\cite{NIST_2013}. After a decade, in 2023, the Ascon family of symmetric authenticated ciphers has been announced as a winner~\cite{NIST_2023}. The algorithm is designed to protect data exchanged between IoT electronics. However, the issue of finding a lightweight and quantum resistant secret key distribution technique has not been yet resolved. As it stands at present, all approved mechanisms rely on the employment of longer keys~\cite{NIST_2022}. This leads to increased complexity and contradicts to the requirements for low latency and low footprint. Therefore, the current state of post-quantum innovations does not align well with the expectations towards 6G networks.

In this sense, PLS-based SKG is a promising quantum-secure approach, that can provide confidentiality at a low computational cost. This is a major advantage compared to standardized solutions. The SKG procedure follows a sequence of steps which allow a pair of devices to extract a shared secret key. A standard source of randomness used to implement SKG is the wireless channel. Reciprocity and stochastic behavior are the properties of the channel that provide the ground SKG~\cite{ZHANG,Mitev_EURASIP2020, MITEV_OJVT_2023}. Devices can extract randomness from their channel observations at a low cost and without the need to perform complex modulo arithmetic in large fields as in current solutions. However, it is important to note that, the achievable key generation rates depend on the statistical properties of the channel, i.e., SKG rates in static channels are significantly lower than in mobile environments. This makes SKG a good candidate for key distribution in many joint communication and sensing applications, where mobility is expected. 

Only a few studies have yet considered the security of joint communication and sensing applications~\cite{secure_JCAS, multicarrier_chirp, Andre_chirp_SKG_2020, Zoli_VTC21}. The work in~\cite{secure_JCAS} provides an information-theoretic perspective on the information leakage between sensing and communication in a discrete memoryless broadcast channel. The considered scenario assumes a transmitter that aims to simultaneously perform channel state estimation and reliable communication with a legitimate user. 
Next,~\cite{multicarrier_chirp} proposes a multi-carrier chirp approach to perform joint communication and sensing. The aim of the study is to maximize the secrecy rate of a wiretap channel in the presence of one or multiple eavesdroppers.
Through numerical evaluation it is demonstrated that 
sensing can provide valuable information about potential eavesdroppers. Finally, the authors of~\cite{Andre_chirp_SKG_2020} and~\cite{Zoli_VTC21} provide simulation-based analyses on the performance of chirp signals for key generation. 
In particular~\cite{Andre_chirp_SKG_2020}, provides an estimate of the bit mismatch probability between two nodes in a vehicular and a smart factory scenarios. The work in~\cite{Zoli_VTC21} provides a different perspective and investigates the number of bits that are leaked to eavesdroppers in the vicinity.

All studies above show promising results proving that employing PLS can be beneficial for future networks. Based on that, we examine a filterbank-based SKG approach, suitable for radar-centric application, where chirp signals are employed. Such scenarios would typically require high sensing precision and low-rate transfer of sensitive information. Multiple use cases that fit this description can be found in the 3GPP report~\cite{3gppSidelink}. A particular example is collective situational awareness, where devices need to simultaneously perform sensing with high precision and exchange small packets at low latency. With a successful history in radar application and the advantage of low-complexity processing  wideband chirp signals are seen as a viable waveform candidate for such scenarios. Furthermore, chirps can  provide a broad view of the wireless channel and offer rich entropy source for SKG. From this perspective, channel measurements (e.g., for equalization purposes) can be used towards SKG and subsequently, secure the information exchange. The idea of the filterbank-based SKG with chirp signals is to divide the received signal bandwidth into a set of parallel sub-bands and measure the power at each sub-band. As compared to RSS-SKG methods, where single value is obtained per transmission, the filterbank approach allows to capture a series of power measurements that represent the randomness of the channel.  


\section{System model} \label{sec:sys_model}

The system model in this work consists of three parties: two legitimate users, namely Alice and Bob, and an eavesdropper, namely Eve. Each of the parties has a single antenna. To measure the channel Alice and Bob exchange a series of linear complex up-chirp signals. Eve is positioned in the vicinity of Bob and eavesdrops their exchange. The goal of Alice and Bob is to generate a secret key using the observed channel measurements while keeping the leakage to Eve close to zero. 
A chirp signal can be represented in baseband as $x(t) =  \frac{1}{T}e^{j \pi c t^2}$, 
where $T$ denotes the duration of the chirp, $c=B/T$ is chirp rate and $B$ denotes bandwidth. The instantaneous frequency changes linearly with time,  $f(t) = ct+f_0$, 
where $f_0=f_c - B/2$ is the starting frequency at $t=0$ and $f_c$ is the center frequency of the signal. 



To perform SKG, Alice and Bob send signals $x(t)$, up converted to a chosen passband. The received signals at Alice, Bob and Eve are: 
\begin{align}
    &y_A(t)=x(t)*h(t)+w_A(t), \label{eq:A-measure}\\
    &y_B(t)=x(t)*h(t)+w_B(t), \label{eq:B-measure}\\
    &y_E(t)=x(t)*h_E(t)+w_E(t), \label{eq:E-measure}
\end{align}
respectively. In the equations above, $h(t)$ is a complex channel response representing the channel between Alice and Bob, $h_E(t)$ represents the channel between Alice and Eve, which, depending on Eve's position may be or may not be correlated to $h(t)$,  
$w_A(t) \sim \mathcal{N}(0,\sigma^2_A)$,  $w_B(t)\sim \mathcal{N}(0,\sigma^2_B)$, $w_E(t)\sim \mathcal{N}(0,\sigma^2_E)$ are additive white Gaussian noise (AWGN) variables experienced at Alice, Bob and Eve respectively.
Please note that this study considers an eavesdropper located close to Bob, hence, measurements in the channel Bob-Eve are ignored.
The channel impulse response is a sum of multi-path components (MPCs), 
\begin{equation} \label{eq:cir}
   h(t) = \sum^{L-1}_{i=0}{ \alpha_i \delta( t-\tau_{i, } ) },
\end{equation}
where $L$ is the number of MPCs, $\alpha_i$ and $\tau_i$  are the complex amplitude and delay of $i$-th MPC, $\delta(t)$ is the Dirac delta function ($h_E(t)$ is defined similarly). Having their channel observations all parties perform filterbank-based SKG as detailed in the next section. 

\section{SKG Protocol} \label{sec:SKG}

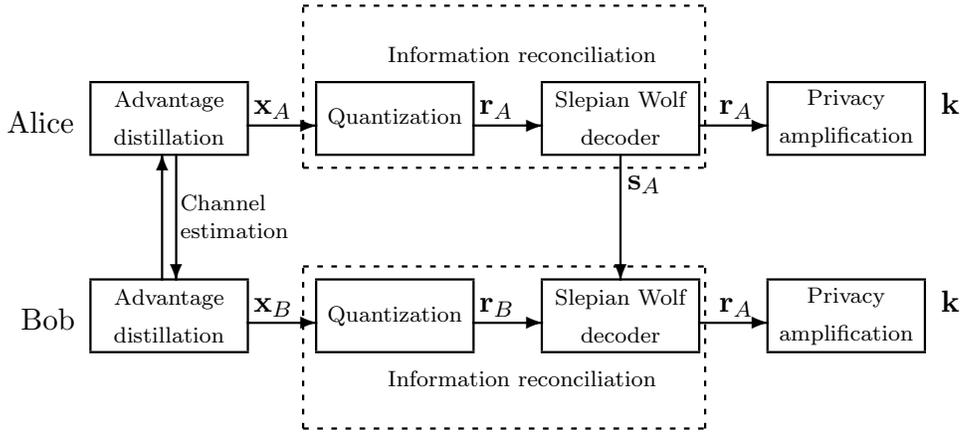
\begin{figure}[!t]
 \resizebox{8cm}{!}{
\setlength{\unitlength}{0.07in} 
\centering
\begin{picture}(40.5,12.5) 
\centering
\thicklines
  \put(0,21.5){{{Alice}}}
 \put(1,7.5){{{Bob}}}
\put(6,6){\framebox(11,5){\Longstack[c]{\scriptsize Advantage \\ \scriptsize distillation }}}
\put(6,20){\framebox(11,5){{\Longstack[c]{\scriptsize Advantage \\ \scriptsize distillation }}}}
\put(11,11){\vector(0,1){9}}
\put(12,20){\vector(0,-1){9}}
\put(12.3,16) {\Longstack[l]{\scriptsize Channel }} 
\put(12.3,14) {\Longstack[l]{\scriptsize estimation}} 
 \put(17,8){\vector(1,0){5}}
\put(17,22){\vector(1,0){5}}
\put(17.5,22.8) {{$\mathbf{x}_{A}$}} 
 \put(17.5,8.8) {{$\mathbf{x}_{B}$}}
\put(22,6){\framebox(11,5){\Longstack[c]{ \scriptsize Quantization }}}
\put(22,20){\framebox(11,5){\Longstack[c]{\scriptsize Quantization }}}
 \put(33,8){\vector(1,0){5}}
\put(33,22){\vector(1,0){5}}
\put(33.5,22.8) {{$\mathbf{r}_{A}$}} 
\put(33.5,8.8) {{$\mathbf{r}_{B}$}} 
\put(38,6){\framebox(11,5){\Longstack[c]{\scriptsize Slepian Wolf\\ \scriptsize decoder }}}
\put(38,20){\framebox(11,5){\Longstack[c]{\scriptsize Slepian Wolf\\ \scriptsize decoder }}}
\multiput(21,19)(1,0){29}
{\line(1,0){0.4}}
\multiput(21,19)(0,1){12}
{\line(0,1){0.4}}
\multiput(21,30.5)(1,0){29}
{\line(1,0){0.4}}
\multiput(49.5,19)(0,1){12}
{\line(0,1){0.4}}
\multiput(21,0.5)(1,0){29}
{\line(1,0){0.4}}
\multiput(21,0.5)(0,1){12}
{\line(0,1){0.4}}
\multiput(21,12)(1,0){29}
{\line(1,0){0.4}}
\multiput(49.5,0.5)(0,1){12}
{\line(0,1){0.4}}
\put(27,26.5) {{\scriptsize Information reconciliation}}
\put(27,3.5) {{\scriptsize Information reconciliation}}
\put(43.5,20){\vector(0,-1){9}}
\put(44,17.5) {{$\mathbf{s}_A$}}
\put(49,8){\vector(1,0){5}}
\put(49,22){\vector(1,0){5}}
\put(50.5,22.8) {{$\mathbf{r}_{A}$}} 
 \put(50.5,8.8) {{$\mathbf{r}_{A}$}}
\put(54,6){\framebox(11,5){\Longstack[c]{\scriptsize Privacy\\ \scriptsize amplification }}}
\put(54,20){\framebox(11,5){\Longstack[c]{\scriptsize Privacy \\ \scriptsize amplification }}}
 \put(65.5,8.8) {{ $\mathbf{k}$}}
 \put(65.5,22.8) {{ $\mathbf{k}$}} 
\end{picture}
 }
 \caption{Secret key generation between Alice and Bob.}\label{fig:Miro_SKG}
\end{figure}

The SKG protocol follows three basic steps, namely, advantage distillation, information reconciliation and privacy amplification. A sketch of the protocol is illustrated in Figure \ref{fig:Miro_SKG}. Alice and Bob pass their channel estimation vectors through a quantizer, obtaining binary sequences $\mathbf{r}_{A}$ and $r_{B}$, respectively. To reconcile mismatches at the generated binary sequences, one of the legitimate parties, e.g. Alice, sends syndrome information $\mathbf{s}_A$ to Bob. Finally, to create maximal entropy secret keys, both users employ privacy amplification over the reconciled information. A detailed description of our implementation of the protocol is given below.

\paragraph*{Advantage distillation:}

\begin{figure}[!t]
    \centering
    \includegraphics[clip, trim=3cm 2cm 3cm 4cm, width=\textwidth]{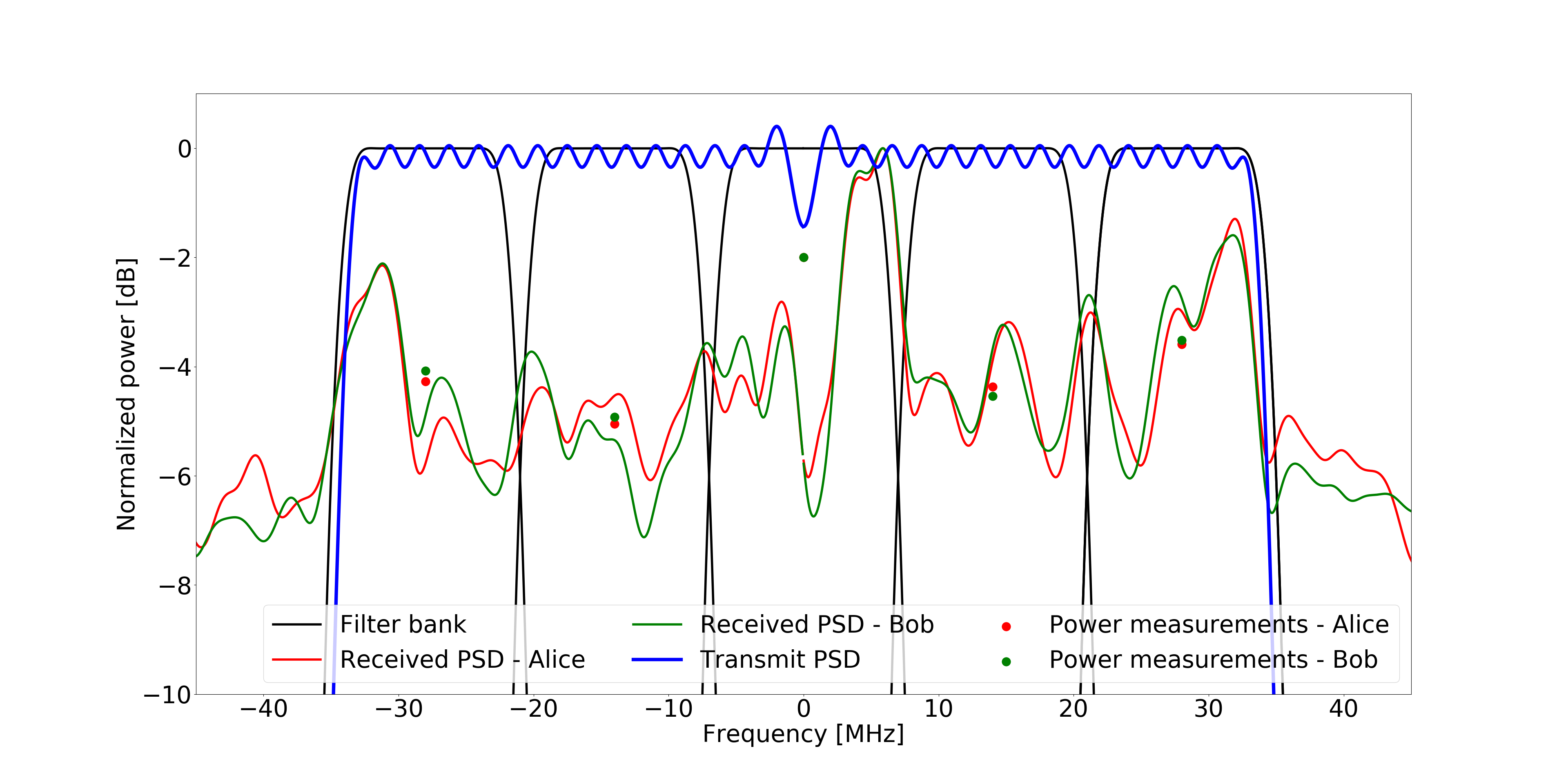}
    \caption{Filterbank constructed of $N=5$ filters. The transmit signal has bandwidth $B=70$ MHz. Power measurements at each sub-band are given as dots.  }
    \label{fig:filterbank}
    \vspace{-0.4cm}
\end{figure}

To extract randomness from \eqref{eq:A-measure}-\eqref{eq:E-measure}, each party passes its received signal through a filterbank with $N$ filters. Each individual filter $n=1, \dots, N$ has an impulse response given as
\begin{equation}
    g_n(t) = g(t)e^{(j 2 \pi f_n(t))},
\end{equation}
where $g(t)$ is a protyping filter. Each filter has bandwidth $B/N$ and center frequency, $f_n(t)$ calculated as $f_n=-\frac{B(N-2n+1)}{2N}$. The randomness from the channel is extracted through power measurements at the output of each filter. An example is illustrated in Figure \ref{fig:filterbank}. The figure shows a normalized power spectrum densities (PSDs) of transmit and received signals with $B=70$ MHz. The impulse response of the filterbank, for $N=5$, is also illustrated along with the power measurements obtained at the output of each filter.
The obtained vector of power measurements at Alice can be expressed as $\mathbf{{p}}_A=\left[{\hat{p}}_{{A,1}} \cdots {\hat{p}}_{{A,N}} \right].$
The measurements at Bob and Eve can be similarly expressed. 
Assuming that measurements are obtained over long enough periods, the power of the sampled version of the filtered signal can be approximated as the average of the continuous-time signals, i.e., ${{\hat{p}}}_{{A,n}}=\overline{|g_{n}(t)*y_A(t)|^2}$ with $\overline{(\cdot)}$ denoting  time-averaging operator. 

The probability distribution of a single power measurement, considering a Rayleigh channel, has been evaluated in our previous work~\cite{MITEV_vtc2022}. It was shown that the probability is a sum of Gamma-distributed variables $\sum_{m=1}^M \Gamma (\kappa_m, \theta_m)$, with shape $\kappa_m$ and scale $\theta_m$ parameters and can be approximated as  
\begin{align}
     \hat{p}_n \sim \Gamma \! \! \left (\! \kappa\!\!=\!\!\frac{(\sum_{m=1}^{M}\kappa_m\theta_m\!+\!2\sigma^2)^2}{\sum_{m=1}^{M}\kappa_m\theta_m^2\!+\!\frac{8\sigma^4}{K-1}}, \theta\!\!=\!\!\frac{\sum_{m=1}^{M}\kappa_m\theta_m \! +\! 2\sigma^2}{\kappa} \!\right)\!\!, \label{eq:power_PDF_alice}
\end{align}
where $M$ is the number of resolvable MPCs, $K$ is the number of digital samples that fall within the filter bandwidth and $\sigma^2$ is AWGN variance. 




\paragraph*{Information reconciliation:}
Power measurements at each party are quantized to bit sequences. Depending on the type of quantizer each power measurement can generate one or multiple bits. Using Gray codes, the number of bits per measurement is $\log_2(Q)$, where the number of quantization levels (regions), $Q$, is chosen to be a power of $2$. After performing quantization the three parties obtain binary vectors $\mathbf{r}_A, \mathbf{r}_B, \mathbf{r}_E \in \{0,1\}^{N\log_2(Q)}$. Due to noise or imperfect power estimation the observations at Alice and Bob could differ.

To correct errors, Alice (or Bob) generates and sends reconciliation information (e.g., syndrome) $\mathbf{s}_A$ over a public channel. The syndrome can be generated using distributed source coding techniques. Next, $\mathbf{s}_A$ is used by Bob to correct $\mathbf{r}_B$ and obtain $\mathbf{r}_A$. However, due to public exchange of the reconciliation information Eve also obtains $\mathbf{s}_A$. She can use the syndrome and her initial measurements to correctly guess part or the entire sequence. To remove leakages to Eve and limit her chances of guessing the legitimate key a final step is performed, i.e., privacy amplification. 



\paragraph*{Privacy amplification:}
Keys are the core of cryptographic mechanisms; hence, their quality is of utmost importance for the security of modern communication systems. In this sense, the final key $\mathbf{k} \in \mathcal{K}$ between Alice and Bob should satisfy the following randomness inequality~\cite{min_entropy} 
$|\mathbf{k}| \leq H_{\infty}(\mathbf{r}_A)$,
where
\begin{equation}
    H_{\infty}(\mathbf{r}_A) =-\log_2\max_{\mathbf{r}_A \in \mathcal{R}_A}p(\mathbf{r}_A), \label{eq:min_entropy}
\end{equation}
denotes min-entropy~\cite{min_entropy-Smith2013}. However, the equation above does not consider any leakage at Eve. Therefore, a more rigorous metric is required to represent the scenario in this work. To measure the quality of generated keys in such situations, conditional min-entropy is used. It is defined as~\cite{cme}: 
\begin{equation}
    H_{\infty}(\mathbf{r}_A|\mathbf{r}_E, \mathbf{s}_A)=-\log_2\max_{\mathbf{r}_A \in \mathcal{R}_A, \mathbf{r}_E \in \mathcal{R}_E, \mathbf{s}_A \in \mathcal{S}_A}p(\mathbf{r}_A|\mathbf{r}_E, \mathbf{s}_A), \label{eq:cme}
\end{equation}
where the shared sequence between Alice and Bob is conditioned on both, initial measurements at Eve and the leaked syndrome. In fact, the difference between  \eqref{eq:min_entropy} and \eqref{eq:cme} represents the total amount of leaked information~\cite{Quant_inf_flow}:
\begin{equation}
    \text{Leakage} = H_{\infty}(\mathbf{r}_A) - H_{\infty}(\mathbf{r}_A|\mathbf{r}_E, \mathbf{s}_A)
\end{equation}
Following the above, the final key should be of size $|\mathbf{k}| \leq H_{\infty}(\mathbf{r}_A|\mathbf{r}_E, \mathbf{s}_A)$.

The evaluation of these quantities requires knowledge on the underlying marginal and conditional distributions. Finding a closed-form solution is a complex task and falls out of the scope for this chapter. Instead, we use a numerical estimator that can evaluate the above without prior knowledge on the analytical expressions. We use the fast black-box leakage estimation (FBLEAU)~\cite{FBLEAU}. To evaluate the leakage and arrive at the conditional min-entropy, the estimator relies on nearest neighbor and frequentist estimates~\cite{leakiEst}. Once the conditional min-entropy is evaluated a standard way to compress the sequence to the desired size is through the use of one-way collision resistant compression functions, e.g., hash function.



\section{Measurement setup}\label{sec:measurement_campaign}

\begin{figure}[!t]
    \centering
    \includegraphics[clip, trim=6.9cm 4.5cm 10cm 1cm, width=0.9\textwidth]{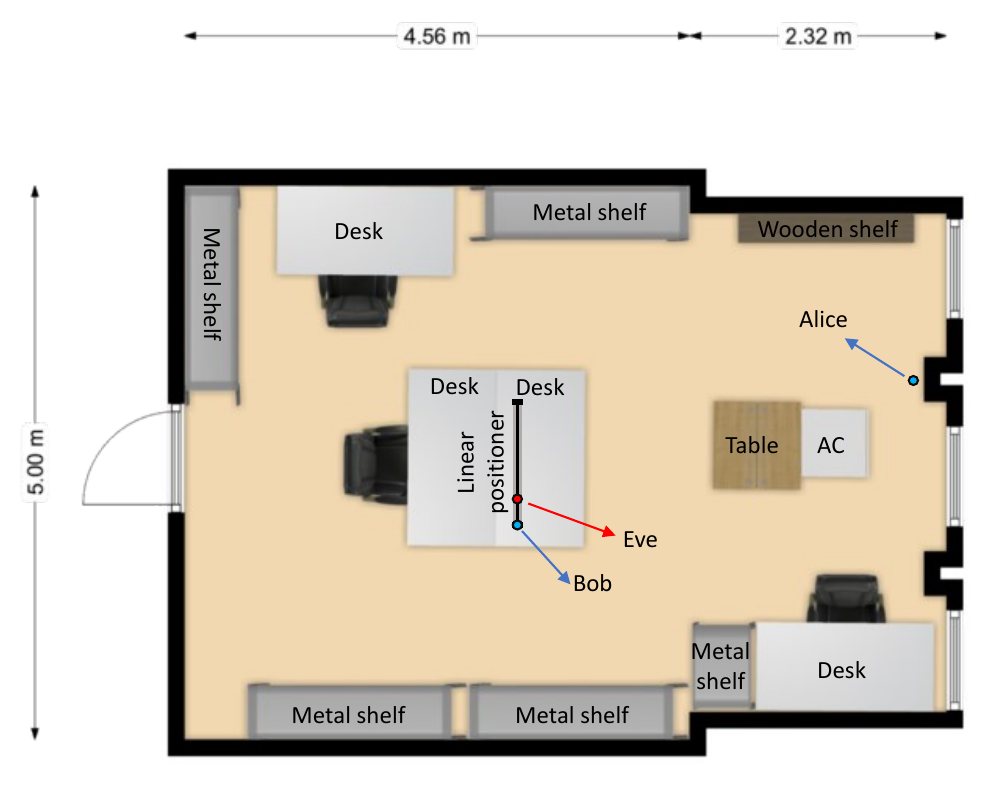}
    \caption{Sketch of the measurement setup created with floorplaner.com. The distance between Alice and Bob is $4$ m. Alice is placed at $2.3$ m height. Bob and Eve are at $0.85$ m height. Throughout the measurement campaign Eve is placed on a linear positioner which moves her with one-wavelength steps with respect to Bob's position (from $1\lambda$ to $10\lambda$).  }
    \label{fig:lab_dimensions}
    \vspace{-0.4cm}
\end{figure}

This section illustrates a real-life setup configured to execute the SKG protocol presented in Section~\ref{sec:SKG}. Based on the system model from Section~\ref{sec:sys_model} three universal software radio peripherals (USRPs), two representing the legitimate nodes (Alice and Bob) and a third acting as an eavesdropper (Eve), were used. For the purpose of this study, USRP-2974 from National Instruments are employed. A floorplan of the measurement setup is illustrated in Figure \ref{fig:lab_dimensions}. The distance between Alice and Bob is $4$ m. Alice is placed at $2.3$ m height, Bob and Eve are placed on a desk at $0.85$ m height. The positions of Alice and Bob are kept constant throughout the measurement campaign. To evaluate the leakage towards Eve, ten different positions are considered for her node. As it can be seen in the figure, she is placed on a linear positioner next to Bob which moves her with steps of one-$\lambda$, starting from $1\lambda$ and ending at $10\lambda$. For the considered frequency, i.e., $f_c=3.75$ GHz, the wavelength is $\lambda \approx 8$ cm. 

To obtain channel measurements Alice and Bob transmit chirp signals in a time division duplex (TDD) fashion. The USRP, configured as Eve, is always in receiving mode and eavesdrops the exchange between the legitimate nodes. Note that, due to her position, Eve is only interested in the signals received from Alice. The duration of individual chirp symbol is $17.1875$ $\mu$s and the bandwidth is $70$ MHz.
At each of the $10$ positions of Eve, Alice and Bob transmit/receive $10^5$ chirp signals. The process described above is repeated considering four different scenarios. 

\begin{figure}[!t]
    \centering
    \includegraphics[clip, trim=0cm 0cm 0cm 0cm, width=\textwidth]{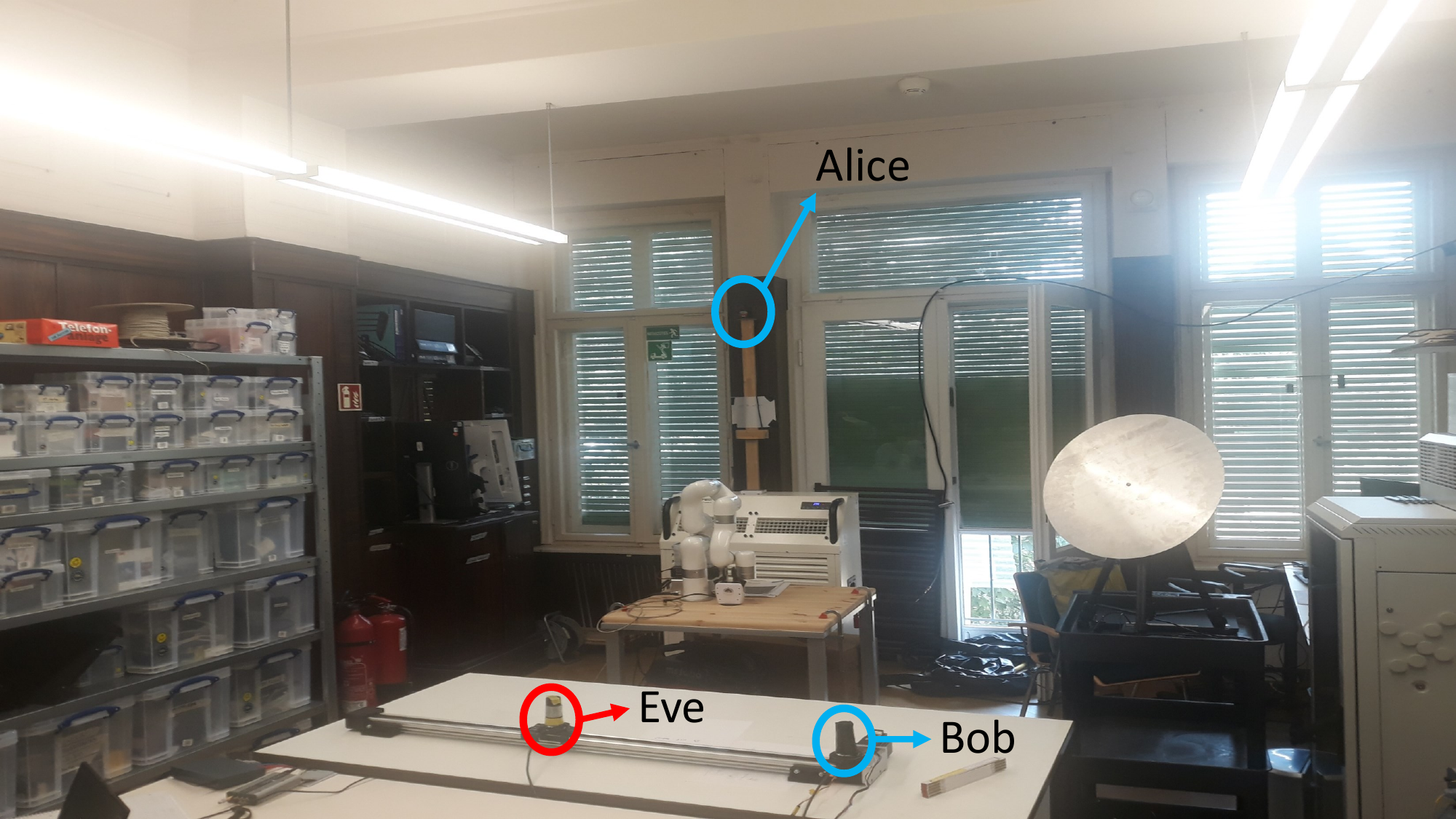}
    \caption{Line-of-sight scenario. Dynamic and static cases differentiate based on whether there is movement in the room. Eve is placed on a linear positioner and is moved from $1\lambda$ up to $10\lambda$ with respect to Bob's position. }
    \label{fig:los}
    \vspace{-0.4cm}
\end{figure}

\begin{figure}[!t]
    \centering
    \includegraphics[clip, trim=0cm 0cm 0cm 0cm, width=\textwidth]{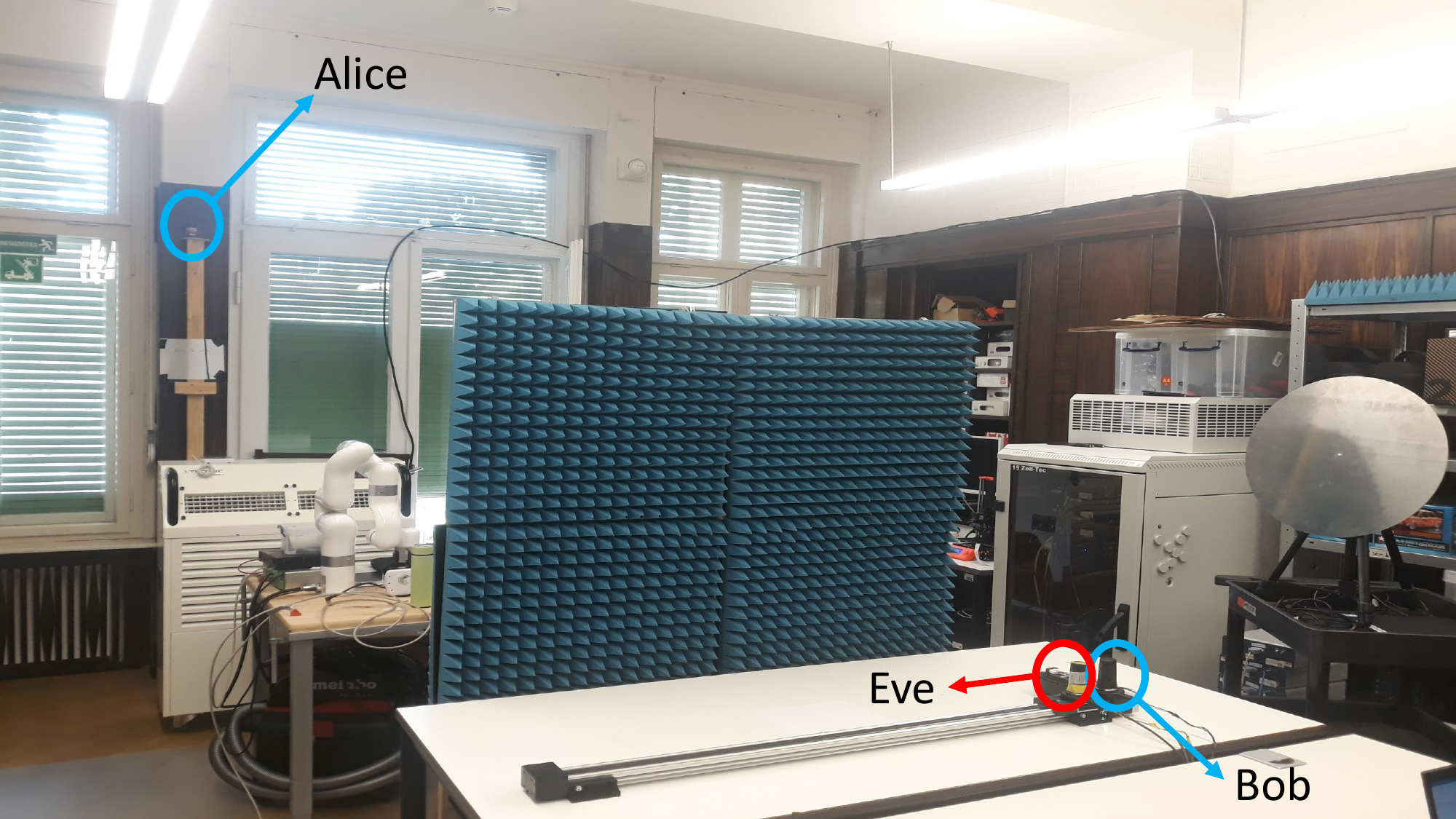}
    \caption{Non-line-of-sight scenario achieved using a set of absorbers to block the direct path between nodes. Dynamic and static cases differentiate based on whether there is movement  in the room. Eve is placed on a linear positioner and is moved from $1\lambda$ up to $10\lambda$ with respect to Bob's position.}
    \label{fig:nlos}
    \vspace{-0.4cm}
\end{figure}

\paragraph*{Scenarios:}
The measurements were performed under a set of four different scenarios including: line-of-sight (LoS) static, LoS dynamic, non-LoS (NLoS) static, NLoS dynamic. Dynamic measurements are achieved by movement of people and objects in the room. Static measurements are performed throughout night and nothing moves during the process. The LoS scenario is illustrated in Figure \ref{fig:los}. One of the objects moved throughout the campaign, i.e., a circular metal plate, is also displayed on the figure. The NLoS case is illustrated in Figure \ref{fig:nlos}. The difference between the two setups is that, several absorbers are placed to block the direct path between antennas. Apart from that the setup is kept identical as in the LoS scenario. 

\paragraph*{Details on the SKG implementation:} Upon receiving a signal, each of the three nodes executes the steps of the SKG protocol presented in Section \ref{sec:SKG}. First, the received signals $y_A, y_B, y_E$ are passed through filterbanks; different number of filters are considered, i.e., $N=\{8,16,32,64\}$. Power measurements at the output of each filter are used as a source for the SKG. Next, the measurements are converted into binary sequences $\mathbf{r}_A, \mathbf{r}_B, \mathbf{r}_E$ using linear multi-level quantization technique. Two different number of quantization levels are employed, i.e., $Q=\{4, 16\}$, resulting in $2$ or $4$ bits per power measurement, respectively. Bit sequences are reconciled using Slepian-Wolf decoders implemented based on Polar (more details on Slepian-Wolf decoding, including Polar codes can be found in~\cite{Mahdi_reconciliation_codes21}). Alice generates a syndrome $\mathbf{s}_A$ and shares it with Bob (this is also observed by Eve). Bob and Eve, use the syndrome along with their observations and aim at reproducing Alice's bit sequence. Depending on the number and distribution of errors the error correction process might be successful or not. Finally, based on the available conditional min-entropy $H_{\infty}(\mathbf{r}_A|\mathbf{r}_E, \mathbf{s}_A)$ Alice and Bob generate a shared secret key $\mathbf{k}$. A thorough performance comparison with respect to the value of $N$, the value of $Q$, and the choice of reconciliation technique is presented in the next section.


\section{Results and discussion}\label{sec:results}
\begin{figure}
\centering
\begin{subfigure}{.49\textwidth}
  \centering
  \includegraphics[clip, trim=0cm 10cm 19cm 0cm, width=\linewidth]{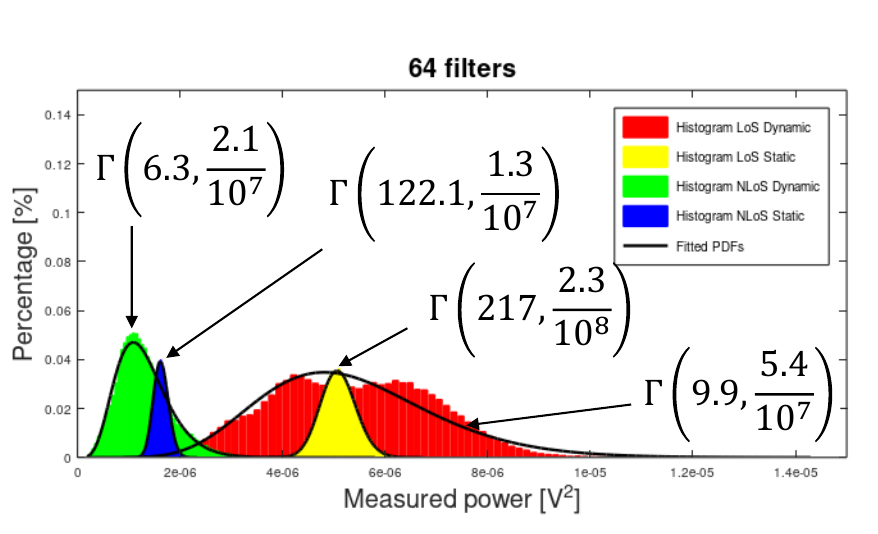}
\end{subfigure}%
\begin{subfigure}{.49\textwidth}
  \centering
  \includegraphics[clip, trim=19cm 10cm 0cm 0cm, width=\linewidth]{Figures/PDFs.pdf}
\end{subfigure}
\begin{subfigure}{.49\textwidth}
  \centering
  \includegraphics[clip, trim=0cm 0cm 19cm 10cm, width=\linewidth]{Figures/PDFs.pdf}
\end{subfigure}
\begin{subfigure}{.49\textwidth}
  \centering
  \includegraphics[clip, trim=19cm 0cm 0cm 10cm, width=\linewidth]{Figures/PDFs.pdf}
\end{subfigure}
    \caption{PDF of the power measurement taken at different environments, all fitted with different Gamma distributions. The measurements are for a single filter when considering $N=64, 32, 16, 8$.}
    \label{fig:PDFs}
    \vspace{-0.4cm}
\end{figure}

This section provides a detailed overview on how the parameters at each step of the SKG protocol  affect the performance. First, we examine our analytical results from~\cite{MITEV_vtc2022}, (also given by Eq. \ref{eq:power_PDF_alice}), and try to fit the histograms of the observed power values with gamma distribution. As noted in the previous section, the transmit signal has a bandwidth of $B=70$ MHz. Filters are distributed evenly across this bandwidth, hence, adding more filters results in smaller bandwidth per filter. This changes the number of digital samples falling within a single filter and has an impact on the power distribution. This is illustrated in Fig.~\ref{fig:PDFs}.


The figure shows that when the channel is static the distribution of the power measurements aligns with our analytical finding, i.e., it fits well with Gamma distribution. Another observation, which was also confirmed in~\cite{MITEV_vtc2022}, is that when the number of filters $N$ decreases (and the bandwidth of each filter increases) the variance of the measurements also decreases.
As a direct consequence of Eq.~\ref{eq:power_PDF_alice} by increasing the number digital samples $K$ (i.e., increasing the bandwidth) the effect of noise $\sigma^2$ decreases. This can be observed for both NLoS and LoS cases. Now, considering dynamic scenarios we can see that only the tail of the distributions align with the Gamma fitted curves and the peaks represent a mixture of several distributions. As expected the variations in dynamic environment are greater from the static ones. However, differently from above, they increase with the bandwidth. This is because the noise is no longer the dominant cause of variation. Instead, multipath provides more degrees of freedom and observing higher bandwidth results in higher spread of the measurements. Overall, this shows that general statistical models cannot fully capture real-life situations, hence, further investigation is required. 

The next step is quantization of the observed measurements. Based on the combination of number of filters and quantization levels different number of bits can be obtained (e.g., for $N=16$ and $Q=4$ we have $16\log_2(4)=32$ bits per channel use). At this step we evaluate the average mismatch probability between nodes. 
\begin{figure}
\centering
\begin{tikzpicture}
\node[inner sep=0pt] at (-0,-15)
{\includegraphics[
clip, trim=47.5cm 23cm 45cm 24cm,
width=0.15\textwidth]{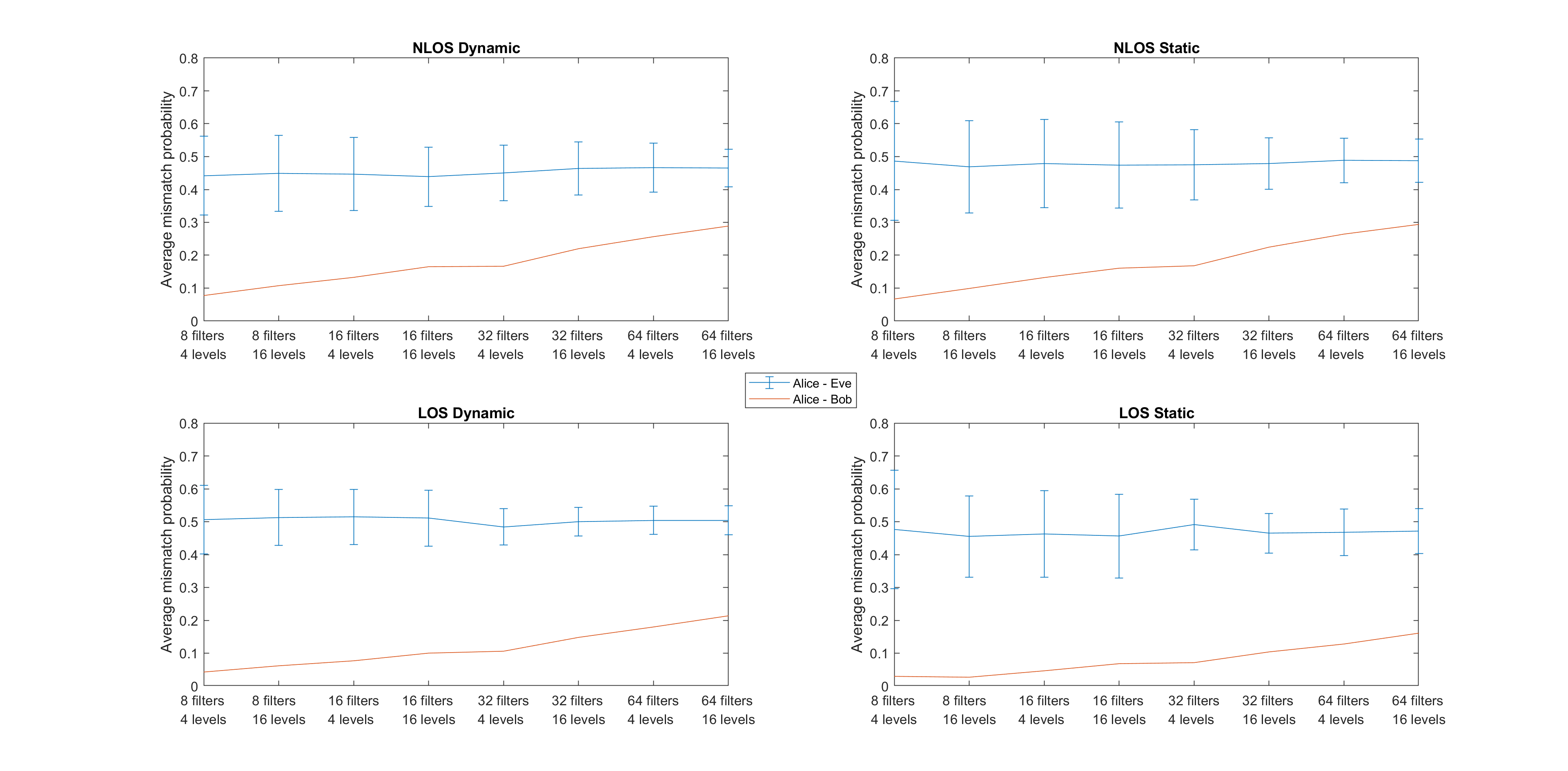}};
\end{tikzpicture} \hspace{5cm}\hfill
\begin{subfigure}{.49\textwidth}
  \centering
  \includegraphics[clip, trim=10cm 26cm 52cm 2cm, width=\linewidth]{Figures/mismatch.png}
\end{subfigure}%
\begin{subfigure}{.49\textwidth}
  \centering
  \includegraphics[clip, trim=54cm 26cm 8cm 2cm, width=\linewidth]{Figures/mismatch.png}
\end{subfigure}
\begin{subfigure}{.49\textwidth}
  \centering
  \includegraphics[clip, trim=10cm 2cm 52.5cm 26.4cm, width=\linewidth]{Figures/mismatch.png}
\end{subfigure}
\begin{subfigure}{.49\textwidth}
  \centering
  \includegraphics[clip, trim=54cm 2cm 8cm 26.4cm, width=\linewidth]{Figures/mismatch.png}
\end{subfigure}
    \caption{PDF of the power measurement taken at different environments, all fitted with different Gamma distributions. The measurements are for a single filter when considering $N=64, 32, 16, 8$.}
    \label{fig:mismatch}
    \vspace{-0.4cm}
\end{figure}

Figure \ref{fig:mismatch} illustrates the average mismatch between legitimate users and the average mismatch observed at Eve. First, as expected, if the legitimate users increase the number of bits generated per frame (i.e., by increasing $N$ or $Q$) the mismatch increases. This is because generated bits become prone to noise variations which are not reciprocal. Second, it can be seen that in the NLoS scenarios the generated bit sequences have more errors as compared to the LoS scenarios. This is expected as the power observed at Alice and Bob decreases when LoS is not present.
Finally, an interesting result can be observed for the mismatch at Eve. The error bars in the figure represent standard deviation of her average mismatch probability taken over all $10$ positions. Note that, during the experiment it was observed that Eve's mismatch does not improve when she is closer to Bob, instead, she can have better probability at further positions - this might be due to in-phase/out-of-phase addition of multi-path components at certain location of the room. This confirms the theoretical result~\cite{half_wavelength}, that channels de-correlate over a distance of $\lambda/2$. The figure shows that, in all scenarios, the average value of Eve's mismatch is independent from $Q$ and $N$ and remains constant close to $50\%$. However, it can be seen that the standard deviation increases for smaller values. This shows that if less bits are produced per frame, i.e., $Q$ and $N$ are small, Eve has higher chances of observing similar bit sequence to the legitimate one.

\begin{figure}
\centering
\begin{tikzpicture}
\node[inner sep=0pt] at (-0,-15)
{\includegraphics[
width=0.65\textwidth]{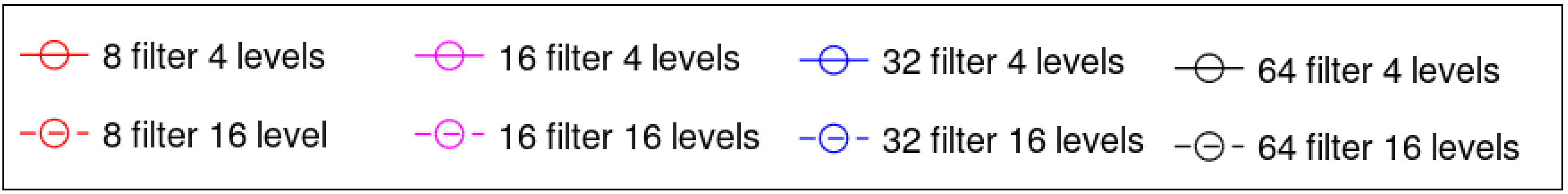}};
\end{tikzpicture} \hspace{5cm}\hfill
\begin{subfigure}{.49\textwidth}
  \centering
  \includegraphics[clip, trim=5cm 15cm 32cm 1cm, width=\linewidth]{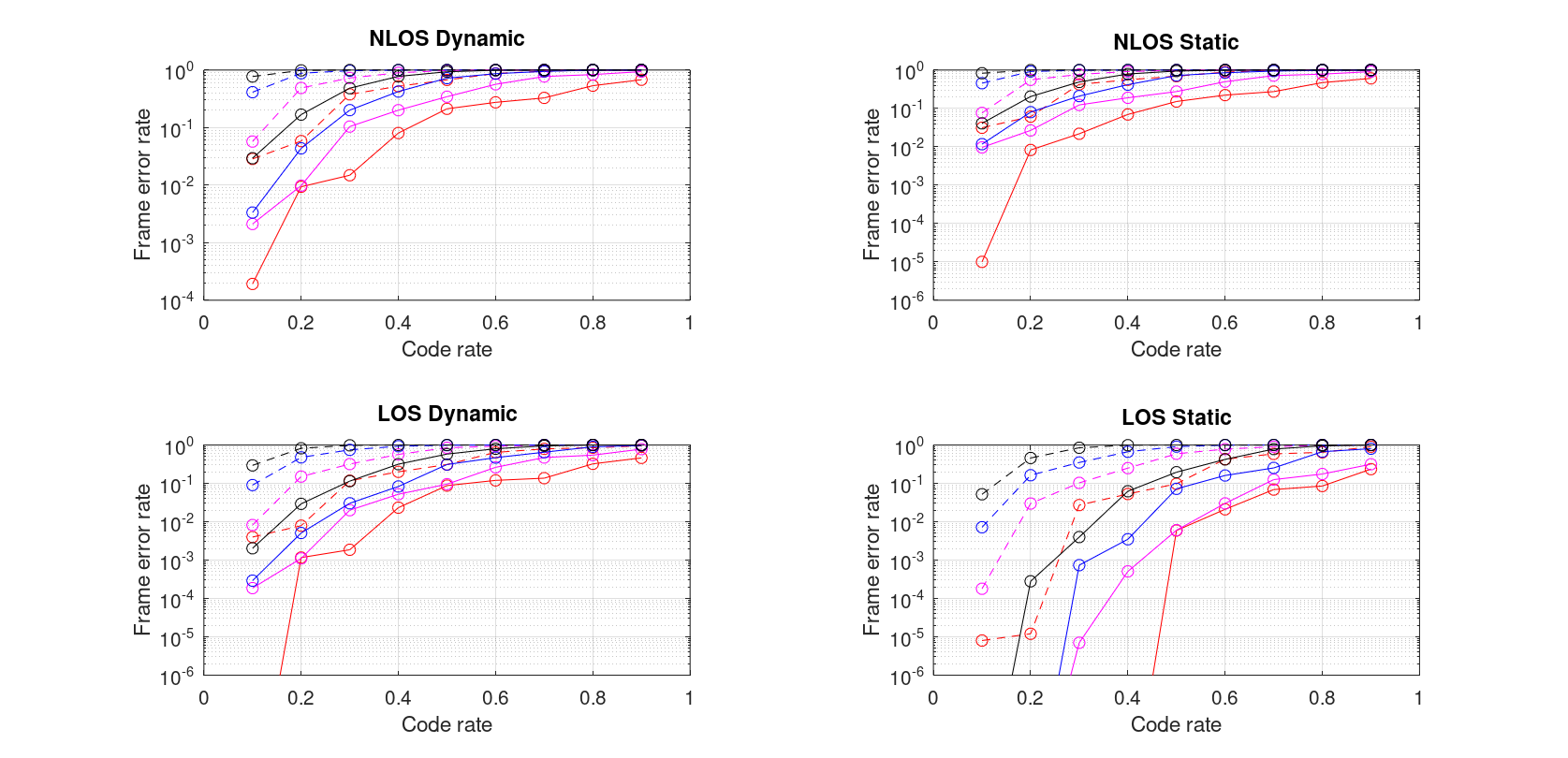}
\end{subfigure}%
\begin{subfigure}{.49\textwidth}
  \centering
  \includegraphics[clip, trim=32cm 15cm 5cm 1cm, width=\linewidth]{Figures/AB_f.png}
\end{subfigure}
\begin{subfigure}{.49\textwidth}
  \centering
  \includegraphics[clip, trim=5cm 1cm 32cm 14cm, width=\linewidth]{Figures/AB_f.png}
\end{subfigure}
\begin{subfigure}{.49\textwidth}
  \centering
  \includegraphics[clip, trim=32cm 1cm 5cm 14cm, width=\linewidth]{Figures/AB_f.png}
\end{subfigure}
    \caption{Frame error rate (FER) at Alice and Bob after reconciliation.}
    \label{fig:fer_AB}
    \vspace{-0.4cm}
\end{figure}

\begin{figure}
\centering
\begin{tikzpicture}
\node[inner sep=0pt] at (-0,-15)
{\includegraphics[
width=0.65\textwidth]{Figures/Capture1.png}};
\end{tikzpicture} \hspace{5cm}\hfill
\begin{subfigure}{.49\textwidth}
  \centering
  \includegraphics[clip, trim=5cm 15cm 32cm 1cm, width=\linewidth]{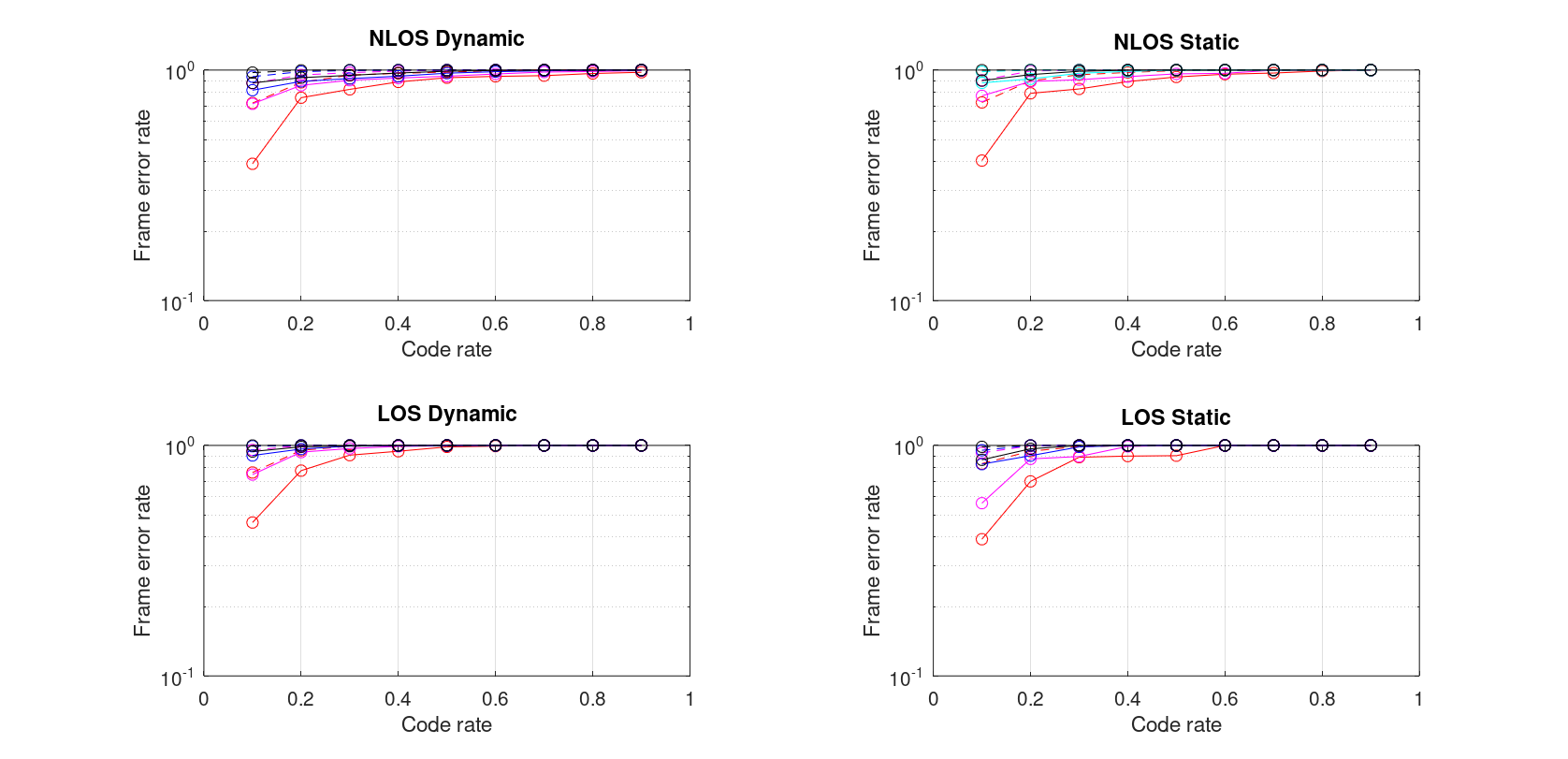}
\end{subfigure}%
\begin{subfigure}{.49\textwidth}
  \centering
  \includegraphics[clip, trim=32cm 15cm 5cm 1cm, width=\linewidth]{Figures/AAA.png}
\end{subfigure}
\begin{subfigure}{.49\textwidth}
  \centering
  \includegraphics[clip, trim=5cm 1cm 32cm 14cm, width=\linewidth]{Figures/AAA.png}
\end{subfigure}
\begin{subfigure}{.49\textwidth}
  \centering
  \includegraphics[clip, trim=32cm 1cm 5cm 14cm, width=\linewidth]{Figures/AAA.png}
\end{subfigure}
    \caption{Frame error rate (FER) at Eve after reconciliation.}
    \label{fig:fer_AE}
    \vspace{-0.4cm}
\end{figure}

Figures \ref{fig:fer_AB} and \ref{fig:fer_AE} illustrate the results from the information reconciliation step. Polar codes as in~\cite{Mahdi_reconciliation_codes21} are implemented to perform this step.

Figure~\ref{fig:fer_AB} shows the frame error rate (FER) at Alice and Bob for different values of $Q$ and $N$. It can be seen that the number of quantization levels $Q$ has a higher impact on the FER as compared to the number of filters $N$. In particular, if Alice and Bob harvest the entropy of the channel using $32$ filter and $4$ ($64$ bits per channel use) levels the FER is smaller compared to $16$ filter and $16$ levels (also $64$ bits per channel use). Similarly to Figure \ref{fig:mismatch}, it is observed that the performance at the legitimate users decreases when a LoS path is not present. This is an expected result, as this naturally decreases the SNR between the two parties. Another important parameter for the success of this step is the code rate. At lower code rate the FER is negligible, however, this corresponds to a larger syndrome $\mathbf{s}_A$. 

The syndrome is sent over a public channel, hence, Eve also has access to it. Therefore, a lower code rate gives her a better chance to arrive at the correct sequence. This is illustrated in Figure \ref{fig:fer_AE}. The figure shows the FER at Eve after she observes the syndrome information. It can be seen that at low code rates, i.e., $0.1$, her probability to decode the correct sequence becomes higher, hence, impacting the security of the system. Important observations from this figure is that, similarly to Alice and Bob, Eve performance becomes better if more filters are used as opposed to more quantization levels. Therefore, while increasing the number of bits by increasing $N$ (instead of $Q$) might be beneficial for the legitimate users, this might increase the potential leakage to Eve. This is examined next.

\begin{figure}
\centering
\begin{tikzpicture}
\node[inner sep=0pt] at (-0,-15)
{\includegraphics[
width=0.95\textwidth]{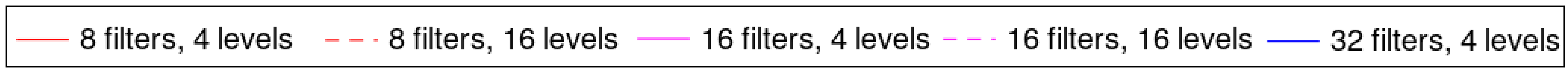}};
\end{tikzpicture} \hspace{5cm}\hfill
\begin{subfigure}{.49\textwidth}
  \centering
  \includegraphics[clip, trim=5cm 15.6cm 33cm 1cm, width=\linewidth]{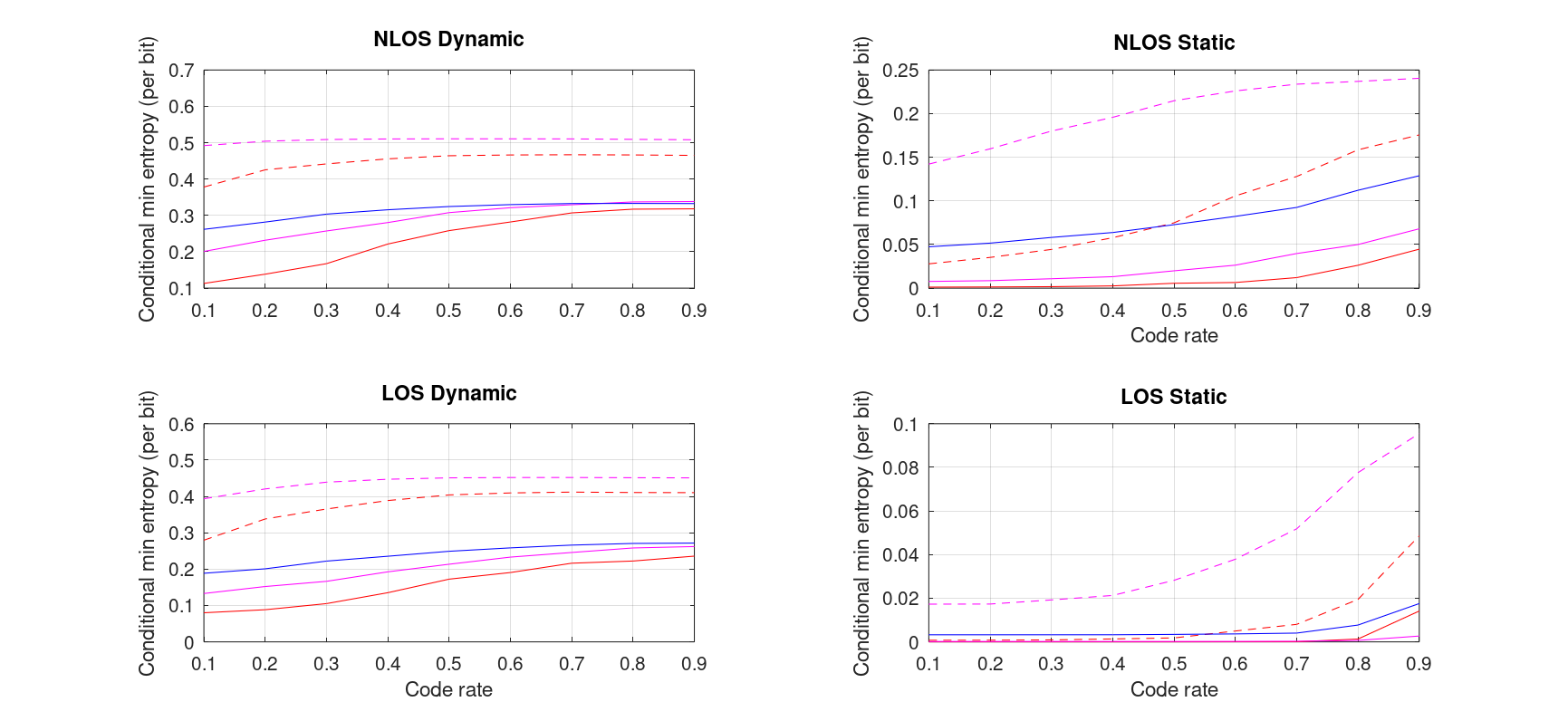}
\end{subfigure}%
\begin{subfigure}{.49\textwidth}
  \centering
  \includegraphics[clip, trim=33cm 15.6cm 5cm 1cm, width=\linewidth]{Figures/cme.png}
\end{subfigure}
\begin{subfigure}{.49\textwidth}
  \centering
  \includegraphics[clip, trim=5cm 0cm 33cm 14cm, width=\linewidth]{Figures/cme.png}
\end{subfigure}
\begin{subfigure}{.49\textwidth}
  \centering
  \includegraphics[clip, trim=33cm 0cm 5cm 14cm, width=\linewidth]{Figures/cme.png}
\end{subfigure}
    \caption{Conditional min entropy  for different $N$, $Q$ and code rates.}
    \label{fig:CME}
    \vspace{-0.4cm}
\end{figure}

We evaluate $H_{\infty}(\mathbf{r}_A|\mathbf{r}_E, \mathbf{s}_A)$ using the FBLEAU numerical estimator~\cite{FBLEAU}. The estimator takes as input the three vectors $\mathbf{r}_A$, $\mathbf{r}_E$ and $\mathbf{s}_A$ and provides an evaluation on the conditional min-entropy. Results are illustrated in Figure \ref{fig:CME}. The figure gives the estimated values for different code rates and different values of $Q$ and $N$. Due to computational constraints, i.e., evaluation over large dimensions of bit sequences, the figure shows only combinations up to $32$ filters with $4$ quantization levels. This limit comes from a problem known as the curse of dimensionality~\cite{CURSE} which is an open research area. It increases the time of convergence non-linearly with each new dimension. 


The first observation is that increasing the code rate increases the conditional min-entropy. This effect is due to the fact that a higher code rate corresponds to a shorter syndrome, hence, less information is leaked to Eve. Furthermore, as the randomness of the generated bits rely on the variation of the channel we can see that the conditional min-entropy is low for static scenarios. In contrast dynamic environments are good source of randomness for Alice and Bob. Finally, controversially to the results for reconciliation we see that capturing the small variations in the channel by increasing $Q$ provides higher randomness as opposed to increasing $N$. However, to execute the full SKG protocol successfully, both reconciliation and conditional min-entropy must be respected. 

\begin{figure}
\centering
\begin{tikzpicture}
\node[inner sep=0pt] at (-0,-15)
{\includegraphics[
width=0.95\textwidth]{Figures/Capture2.png}};
\end{tikzpicture} \hspace{5cm}\hfill
\begin{subfigure}{.49\textwidth}
  \centering
  \includegraphics[clip, trim=5cm 15.6cm 33cm 1cm, width=\linewidth]{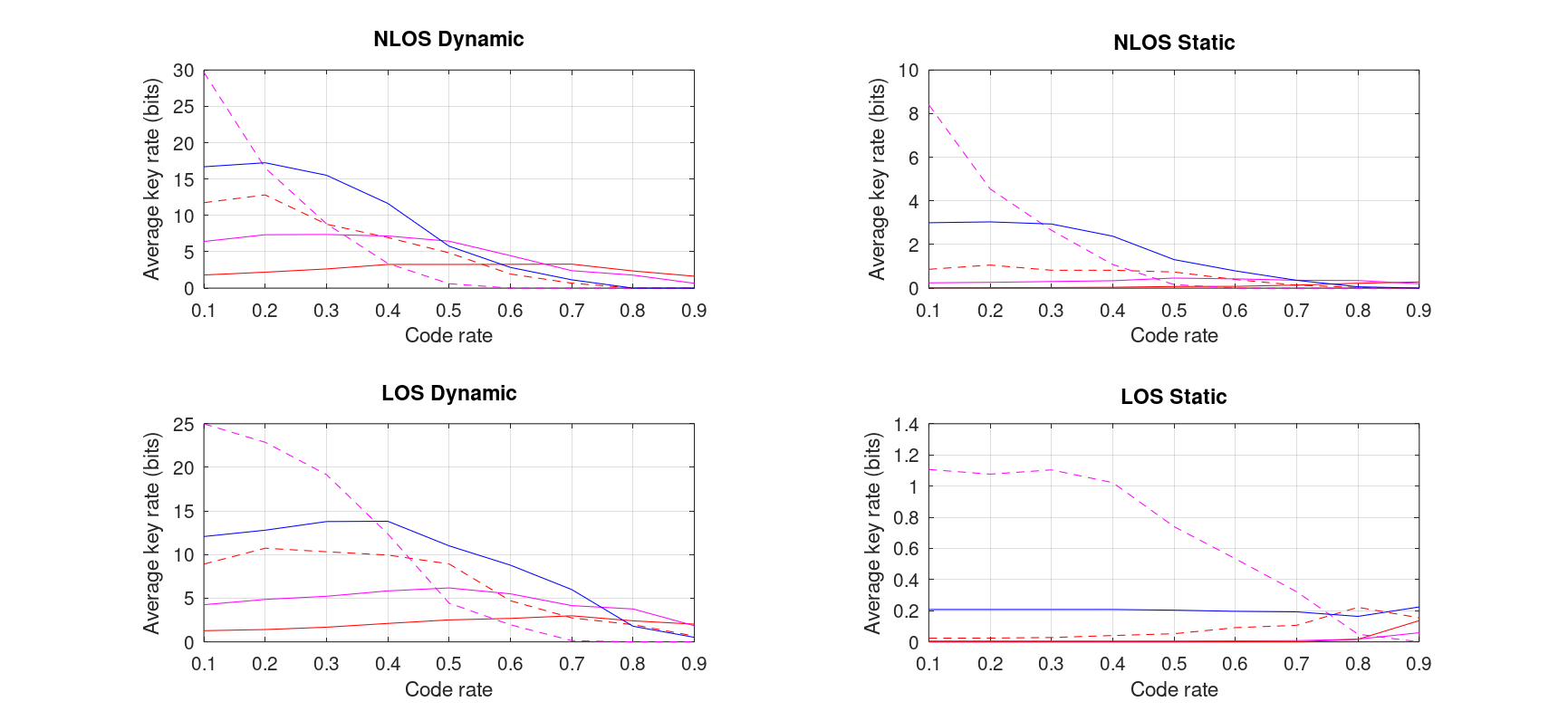}
\end{subfigure}%
\begin{subfigure}{.49\textwidth}
  \centering
  \includegraphics[clip, trim=33cm 15.6cm 5cm 1cm, width=\linewidth]{Figures/key_rate.png}
\end{subfigure}
\begin{subfigure}{.49\textwidth}
  \centering
  \includegraphics[clip, trim=5cm 0cm 33cm 14cm, width=\linewidth]{Figures/key_rate.png}
\end{subfigure}
\begin{subfigure}{.49\textwidth}
  \centering
  \includegraphics[clip, trim=33cm 0cm 5cm 14cm, width=\linewidth]{Figures/key_rate.png}
\end{subfigure}
    \caption{Average key rate (per channel use) over different $N$, $Q$ and code rate.}
    \label{fig:final_rate}
    \vspace{-0.4cm}
\end{figure}

The average key rate per channel use, $R$ per channel use is a function of several parameters: i) frame rate $F$, equal to the number of bits generated per frame (varies accordingly with $Q$ and $N$); ii) the FER after reconciliation; iii) conditional min-entropy $H_{\infty}(\mathbf{r}_A|\mathbf{r}_E, \mathbf{s}_A)$. The key rate is denoted as:
\begin{equation}
    R= F \times (1-\text{FER}) \times H_{\infty}(\mathbf{r}_A|\mathbf{r}_E, \mathbf{s}_A).
\end{equation}
Results on the rate are given in Figure \ref{fig:final_rate}. 

The trend in the figure is clear. To increase the key rate, $F$ must be increased, preferably by increasing $Q$ as opposed to $N$ and the code rate should be decreased. 
Based on the system parameters used in our measurement campaign the values from Fig. \ref{fig:final_rate} can also be expressed in $[$b/s/Hz$]$ or $[$b/s$]$. Given the fact that a two-way transmission between Alice and Bob takes approximately $35$ $\mu$s and $B=70$ MHz we present the results in Table \ref{tab:rates}. The table shows only the maximum achievable rate based on Fig. \ref{fig:final_rate}, i.e., considering code rate $0.1$, $N=Q=16$. 

\begin{longtable}[c]{| l | c | c | c | c |}
 \caption{Achievable secret key generation rates.\label{tab:rates}}\\
 \hline
 Scenario & NLoS Dynamic & NLoS Static & LoS Dynamic & LoS Static \\
 \hline
Rate [b/s/Hz] & 0.0121 & 0.0034 & 0.0102 & 0.0005 \\ \hline
Rate [b/s] & 848220 & 240131 & 713940 & 31614 \\
 \hline
 \endfirsthead

 \end{longtable}

To conclude, our experimental results show that the information available for SKG could vary greatly depending on both channel and system parameters. We have shown that chirp signals can capture the channel variations over different frequencies and provide a valid source for SKG. 
The evolution of wireless systems and the introduction of sensing at the edge of the system can provide valuable information regarding the environment. This provides the required context information and makes SKG a viable technique for key distribution. As a future work, the authors plan to further investigate the results presented in this chapter, by considering multiple antenna scenarios and work on adaptive SKG solutions. Furthermore, the proposed chirp-based approach will be experimentally compared with waveforms with high spectral efficiency waveforms (e.g., OFDM).



\section*{Acknowledgement}
This work is financed on the basis of the budget passed by the Saxon State Parliament, and partially funded by Hexa-X-II project, receiving funding from the Smart Networks and Services Joint Undertaking (SNS JU) under the European Union’s Horizon Europe research and innovation programme under Grant Agreement No 101095759. A. Chorti was supported by INEX funding (Project PHEBE and ENIGMA). The authors would like to acknowledge the help from Dr. Mahdi Shakiba-Herfeh who provided us with a Slepian-Wolf implementation of Polar codes.

\backmatter

\bibliographystyle{unsrtnat}
\bibliography{Wiley}%

\end{document}